# Evidence for OH or H$_2$O on the surface of 433 Eros and 1036 Ganymed

Andrew S. Rivkin[1], Ellen S. Howell[2], Joshua P. Emery[3], Jessica Sunshine[4]

1. Johns Hopkins University Applied Physics Laboratory
   11101 Johns Hopkins Rd
   Laurel MD, 20723 USA
   443-778-2811
   andy.rivkin@jhuapl.edu
2. University of Arizona, Lunar and Planetary Laboratory
3. University of Tennessee
4. University of Maryland

*Abstract*

Water and hydroxyl, once thought to be found only in the primitive airless bodies that formed beyond roughly 2.5-3 AU, have recently been detected on the Moon and Vesta, which both have surfaces dominated by evolved, non-primitive compositions. In both these cases, the water/OH is thought to be exogenic, either brought in via impacts with comets or hydrated asteroids or created via solar wind interactions with silicates in the regolith or both. Such exogenic processes should also be occurring on other airless body surfaces. To test this hypothesis, we used the NASA Infrared Telescope Facility (IRTF) to measure reflectance spectra (2.0 to 4.1 μm) of two large near-Earth asteroids (NEAs) with compositions generally interpreted as anhydrous: 433 Eros and 1036 Ganymed. OH is detected on both of these bodies in the form of absorption features near 3 μm. The spectra contain a component of thermal emission at longer wavelengths, from which we estimate thermal of 167±98 J m$^{-2}$s$^{-1/2}$K$^{-1}$ for Eros (consistent with previous estimates) and 214±80 J m$^{-2}$s$^{-1/2}$K$^{-1}$ for Ganymed, the first reported measurement of thermal inertia for this object. These observations demonstrate that processes responsible for water/OH creation on large airless bodies also act on much smaller bodies.

*1 Background*

The near-Earth objects (NEOs), including the subset of non-cometary near-Earth asteroids (NEAs), have a mix of different compositions (Stuart and Binzel 2004, Binzel et al. 2004). The meteorites seen to most commonly fall to Earth are the ordinary chondrites (OCs), derived from undifferentiated bodies whose reflectance spectra in the visible and near-IR (0.4-2.5 μm) are dominated by silicates (Burbine et al. 2002). These spectral characteristics are shared by asteroids in the "S complex" in currently used taxonomies (Bus and Binzel 2002, DeMeo et al. 2009), with the Q class (a subgroup within the S complex) being close spectral matches to the OCs. Some bodies with spectra belonging in the S complex (including many in the S class within the S complex) are thought to be analogs to OCs with more mature regolith, while others are thought to represent igneous meteorites like the stony-irons or primitive achondrites. The meteorites associated with the S complex asteroids do not generally have abundant phyllosilicates or other minerals with structural water or hydroxyl (hereafter called "hydrated minerals"), though there are a handful of exceptions (Hutchison et al. 1987, Zolensky et al. 2000).

Carbonaceous chondrite meteorites, conversely, can have abundant hydrated minerals, with the CM group having the equivalent of 10% $H_2O$ by weight (Jarosewich 1990). These meteorites are generally associated with asteroids in the C complex, which are very common throughout the main asteroid belt but relatively less common in the NEO population. It has been thought that any NEOs with hydrated minerals would be from this group, and several with absorptions indicating hydrated minerals have been found: 1992 UY4 (Volquardsen et al. 2007) and 1996 FG3 (Rivkin et al. 2013), which have an absorption near 3 μm attributed to hydroxyl, and 2002 DH2 (Binzel et al. 2004) and 2001 TX16 (Yang et al. 2003), which have absorptions near 0.7 μm correlated with hydrated minerals.

Previous expectations of the role of water and OH in NEOs were upended by the discovery of OH in the lunar regolith by three spacecraft (Sunshine et al. 2009, Pieters et al. 2009, Clark 2009). The lunar datasets used the 3-µm spectral region, where the hydroxyl fundamental absorption and the very strong first overtone of water are both found.  At least some of the lunar OH has been attributed to interactions between solar wind hydrogen and a nominally anhydrous regolith. Sunshine et al. (2009) pointed out that similar processes should be occurring on asteroids, particularly the NEOs that spend significant times near 1 AU.  Sunshine et al. (2009) also found variable band depths for the same location at the two Deep Impact lunar encounters, the cause of which is still under investigation.

In addition to the lunar observations and their interpretation, there is evidence for an absorption on Vesta due to OH (Hasegawa et al. 2004, De Sanctis et al. 2012). This absorption is reported to be deeper in the older low albedo areas of Vesta, which also is reported to have higher hydrogen abundances in neutron detector data (Prettyman et al. 2012).  This correlation has led to an interpretation, first suggested by Hasegawa et al. (2004) and Rivkin et al. (2006), that the regolith of Vesta is contaminated with the remains of carbonaceous impactors (consistent with the presence of CM xenoliths in HED meteorites: Buchanan et al 1993, Gounelle et al. 2003, and others), and that the hydrated minerals detected were brought in via those impactors (Reddy et al. 2012).  Unlike the Moon, Vesta appears to not have experienced significant OH creation from the solar wind, at least compared to the impactor contribution (Prettyman et al. 2012).

These factors combine to make observations of NEOs in the 3-µm region relevant for not only small bodies studies but also for lunar studies. Those NEOs considered analogs of anhydrous meteorites, particularly the S asteroids, will be least susceptible to confusion between native hydrated material and either exogenic material or OH created through solar wind interactions.

*2 Observations*
The two largest NEOs, 1036 Ganymed and 433 Eros, are excellent matches to the kind of objects desired and described above. Both are in the S class, and from previous observations we do not expect significant amounts of indigenous water on either one of them. Ganymed is interpreted by Fieber-Beyer et al. (2011) as having an orthopyroxene composition for its silicates, perhaps consistent with being part of a mesosiderite parent, or alternately unrepresented in the meteorite collection. Eros is linked to the L or LL chondrites via in-situ data from the NEAR Shoemaker spacecraft (Trombka et al. 2000, Foley et al. 2006, although we note that one hydrated LL chondrite does exist: see below). In the time since the observations reported here were conducted, Peplowski et al. (2015) used gamma-ray data from Eros to measure the water content on that body, allowing us to further calibrate our observations (Section 4.5).

Both Ganymed and Eros made very good apparitions in the 2011-2013 period: both the phase angle and solar distance for Ganymed varied significantly during the apparition, and Eros had a limited range of phase angles while its solar distance varied. This geometry provided the opportunity to separately consider phase angle and solar distance/temperature effects.

We used the NASA Infrared Telescope Facility (IRTF) SpeX instrument to observe Eros and Ganymed. We used SpeX (Rayner et al. 2003) in its long-wavelength cross-dispersed (LXD) mode, simultaneously covering 1.9-4.2 µm. SpeX was set with a 15" N-S beam switch. The observing as well as the data reduction process is the same as for other observations done as part of the L-band Main-belt/Near-Earth object Observing Program (LMNOP), detailed in several earlier papers (Rivkin et al. 2006, Rivkin et al. 2011, Rivkin et al. 2013, etc.). Table 1 shows the observing dates, the standard stars used on each date, which of the target asteroids were observed, and the average precipitable water found through the data reduction (see next section). Tables 2 and 3 focus on Eros and Ganymed specifically, including their observational circumstances.

Thermal flux from the Earth's atmosphere typically limits integration times to 20-25 seconds or less before the longest wavelengths enter the non-linear response regime for SpeX in LXD mode. In addition, atmospheric stability often only lasts on a timescale of a few minutes, with poorer atmospheric subtraction resulting when beam switching on longer timescales. As a result, we used integration times of 10-20 seconds with 1-4 coadds in order to do beam switches every 10-90 seconds. The details of integration and total observing times per night are provided in Table 1.

*3 Data reduction and thermal modeling*
*3.1 General Procedure*
Data reduction with SpeX follows a well-established pipeline, with the IRTF-provided Spextool IDL package (Cushing et al. 2003) used for flat fielding, wavelength calibration, and spectral extraction and combination. Following this, an additional stage is also carried out, with a model atmosphere fit to the data and the contribution of terrestrial absorptions removed. This stage uses additional IDL code developed by S. J. Bus and E. L. Volquardsen of the IRTF, and relaxes otherwise stringent airmass match requirements, allowing a larger set of standard stars to be appropriately used for a given asteroid. The precipitable water fit by this second stage of reduction compares well to the independent measures of optical depth due to water in the Earth's atmosphere made by the Caltech Submillimeter Observatory (CSO), and we include the fit values in Table 1.

Eros and Ganymed were warm enough during the observations that their own thermal emission was a significant fraction of the measured flux, especially at the longest wavelengths (Figure 1). We used a modified version of the Standard Thermal Model (Lebofsky et al. 1986), as we have in previous work (Rivkin et al. 2006, Rivkin et al. 2013, etc.). The inputs to the model are all well known for the targets, save for the emissivity (typically set to 0.9), and the beaming parameter ($\eta$), which serves as a catchall factor for thermal inertia, shape, and other factors that cannot be explicitly considered in this simple model. While not an output of the model per se, we also calculated the sub-solar temperature ($T_{ss}$) for Eros and Ganymed at the time of observation, using the appropriate value for $\eta$ to match the chosen continuum values (see below). These values are also included in Tables 4 and 5. Different values of $\eta$ change the calculated $T_{ss}$, and a change in $\eta$ of 0.1 can change the temperature by ~10 K for a particular observation when everything else is fixed. Energy balance considerations, along with specific values for Bond Albedo and $\eta$, can provide an estimate of average temperature for an asteroid ($T_{av}$), which for Eros and Ganymed is roughly 0.65-0.7$T_{ss}$.

*3.2 Choosing a continuum*
For the majority of asteroids observed in the 3-µm region, typically low-albedo objects in the main belt, the choice of a continuum is relatively straightforward. The spectral region of most interest near 2.9-3.1 µm is not strongly affected by thermal emission at main-belt asteroid temperatures, and continuum behavior is fairly well established shortward of 2.5 µm in carbonaceous chondrites, considered the likely analog for those asteroids (Burbine et al. 2002, Section 1). Laboratory spectra of analog materials in the appropriate wavelength region are available, including those optimized for comparison to asteroids by removal of telluric water. In the case of the C-class asteroid 1996 FG3, which was observed near 1.0 AU and had prodigious thermal flux to be removed, these meteorite and main-belt asteroid observations provided the framework for constraining the continuum and allowing the interpretation of the reflectance spectrum (Rivkin et al. 2013).

For Eros, we used measurements of its 3.6 µm/0.55 µm reflectance ratio reported in Trilling et al. (2010) and the behavior of laboratory measurements of L and LL chondrites to fix the continuum value at 3.8 µm. These independent approaches all result in a consistent expected straight-line continuum with $R_{3.8}/R_{2.45} \sim 1.05$.

There is less consistency when considering the continuum for Ganymed. The Wide-field Infrared Survey Explorer (WISE) mission measured the infrared fluxes for Ganymed, providing a means of estimating the proper continuum for that object. Mainzer et al. (2011) set the 3.37 and 4.62 µm albedos (i.e. those measured through the two shortest-wavelength filters on WISE) equal to one another, denoted as $p_{IR}$, and measured the $p_{IR}/p_V$ ratio as 1.45. The LXD-mode Ganymed data collected in 2011 and presented here cannot simultaneously satisfy $p_{3.37} = p_{4.62}$ and $p_{IR}/p_V = 1.45$, so we choose to set $p_{IR}$ equal to the albedo halfway between the two IR wavelengths, at 4.0 µm. Setting $p_{3.37}=p_{4.62}$ would imply $p_{IR}/p_V \sim 1.25$-1.3, rather lower than the WISE measurements suggest. Unfortunately, meteorite analogs for Ganymed are not as widely agreed-upon as for Eros (see above), and so are of limited use in determining its continuum. The band depths interpreted for Ganymed on particular nights are dependent upon the continuum behavior chosen, but the highest-quality data indicates an absorption band regardless of continuum calculation. For Ganymed, we adopt $R_{3.8}/R_{2.45} \sim 1.10$.

For both asteroids the value of $R_{3.8}/R_{2.45}$ is used to select the correct thermal model. The continuum is defined as a straight line from 2.45-3.4 µm, from which absorptions at 2.95 and 3.10 µm are measured. In this work, the band depth is defined as the difference in reflectance between the measured spectrum and the calculated continuum, using the weighted average value of the reflectance for points within ±0.01 µm of the wavelength of interest.

*3.3 Latitudes and Longitudes*
Thanks to the visit of the NEAR Shoemaker spacecraft, Eros has a well-determined planetographic coordinate system. These latitudes and longitudes, available in JPL's Horizons on-line ephemeris system, are used for Eros in this work and are listed in Table 2. No such coordinate system has been established for Ganymed. However, its rotation period is precisely determined and given the known geometry the relative central longitudes of observation can be calculated, and are presented in Table 3. A pole position for Ganymed is also available (Ďurech et al. 2010), and can be used to calculate planetocentric latitudes, again shown in Table 3. We note that the planetographic coordinates for Eros and its irregular shape can lead to some unexpected lighting conditions: for instance, Table 3 shows that the 6 December 2011 observations had sub-solar and sub-Earth latitudes separated by nearly 150°, though Eros was not between the Sun and Earth at the time. Similarly, the planetographic coordinates for 30 October 2011 suggest a nearly pole-on view, but planetocentric coordinates show that Earth was much closer to the equator (Section 4.2) However, using planetocentric coordinates for Eros does not change any of the conclusions about trends in spectral properties reported below. The planetocentric latitudes used here for Ganymed vary more regularly and expectedly than Eros' planetographic coordinates, but are not sensitive to, and do not reflect, its shape.

*4 Results and Discussion*
*4.1 Beaming parameter and Thermal Inertia*
The values for η give a sense of the thermal inertia of Ganymed and Eros. Delbo et al. (2007) simulated η for model objects with thermal inertias of either 15, 200, or 1000 (in SI units of J m$^{-2}$ s$^{-0.5}$ K$^{-1}$) and observed at a variety of phase angles. We find that both Eros and Ganymed are most similar to the model population with thermal inertia of 200. This is consistent with independent measurements of Eros' thermal inertia (Delbo et al. 2007), and provides the first thermal inertia estimate for Ganymed. We also calculated thermal inertias for each observation using the approach of Harris and Drube (2016), who provided a means of estimating thermal inertia from values of η and the solar aspect angle (which is the same as the sub-solar colatitude). The mean estimate for the thermal inertia of Ganymed in SI units is 214 +/- 80. The estimates for Eros' thermal inertia range from ~80-350 in the same units, with a single estimate of 1000. This last estimate is for a nearly pole-on view of Eros and this estimation technique is likely less valid in that geometry.

Excluding that measurement results in a mean thermal inertia for Eros of 167 +/- 98. These are, again, consistent with previous estimates.

*4.2 433 Eros*
The 25 January 2012 observations of Eros have a happy combination of the asteroid at its brightest and a very low precipitable water value, resulting in the highest S/N of the dataset. Figure 2 shows this spectrum which clearly shows a band depth at 2.95 µm of ~2%, along with the other spectra of Eros. The band shape, though shallow, is reminiscent of what is commonly seen in Ch-class asteroids in the 3-µm region, often called "check-shaped" or "Pallas-type" (Takir and Emery 2012, Rivkin 2010).

The similarity to Ch-type asteroids leads to a possible interpretation of the absorption band as due to carbonaceous impactors, as is the absorption band seen on Vesta (Reddy et al. 2012). The CM, CR, and CI carbonaceous chondrites and Ch asteroids have band depths of roughly 10-30% in the 3-µm region due to hydrated minerals. If the native material has an intrinsic band depth of zero, and the material is mixed in a linear fashion (where a given photon is likely to only encounter one material rather than both hydrated and anhydrous material), the band depth would indicate a contaminator fraction of as much as 10-15%. This is somewhat larger than might be expected, but it is also an upper limit, since a smaller fraction of contaminant is required to provide a similar spectrum in a non-linear intimate mixture. It is also smaller than but consistent with the carbonaceous impactor fraction reported for Vesta by Palomba et al. (2014), which reaches upward of 30% in low-albedo units and is estimated as up to 10% even in the higher albedo units. We also note, though, that NEAR Shoemaker orbital imagery shows no evidence of large, low-albedo areas as are seen on Vesta.

It is also possible the hydrated minerals detected are native to Eros. This is more difficult to test at this point, since spectra of Semarkona, the one OC meteorite with significant amounts of phyllosilicates (Hutchison et al. 1987), are not available in the literature. Such spectra could help establish whether the hydration band seen on Eros requires a significant exogenic contribution or whether it can be explained by native material alone.

The observations at other epochs are at least consistent with those from January. The band depths are roughly 2-5% in most observations, though it is consistent with zero in one observation and very shallow in another. The observations from 30 October differ in their spectra: the set taken earlier in the night has no absorption, while that taken with a mid-time 45 minutes later shows an absorption of ~3% at 2.95 µm. The observing conditions did not change during this period, nor was one set of observations taken at a significantly higher airmass. The best explanation appears to be that of spectral variability on Eros' surface: the viewing angle for the first observation is nearly aligned with the long axis of Eros while the second observation is of the Psyche hemisphere. This is demonstrated in Figure 3, which shows the view of Eros from Earth at the midtime of each observation. Also included in Figure 3 for convenience is the solar distance and 2.95- and 3.1-µm band depths.

It is also evident from Figure 2 that the band centers differ on different nights, falling closer to 3.1 µm on 27 August 2011 and 3 July 2012, though the data quality on the latter date is not high. This could indicate variation in hydrated mineralogy across Eros' surface. An absorption band centered near 3.1 µm is seen on Ceres and has been attributed to various minerals through the years, with the most recent assignment that of ammoniated phyllosilicates (De Sanctis et al., 2015). It has also been seen on Vesta, where a specific assignment has not been offered (De Sanctis et al. 2012). However, Applin et al. (2016) suggest oxalate minerals could explain this band on both Ceres and Vesta. Identifying the specific mineral responsible for this band on Eros will require additional modeling work beyond the scope of this paper.

There are other indications that Eros' surface is heterogeneous in terms of hydrated minerals. While data quality varies, it appears from Figure 3 and Figure 4 that the Eros observations with the smallest band depths were centered near Eros' prime meridian and observations centered on Psyche crater seem to have a deeper band depth at 2.95 µm than 3.1 µm, while those with deeper 3.1 µm band depths seem to be centered on neither longitude. Disentangling the effects of changing solar and observer sub-longitudes and the contributions made to the spectrum by differently lit regions on Eros' surface and how those indicate or constrain hydroxyl on Eros' surface will require a much larger dataset (particularly given the shallow band depths) and is beyond the scope of this paper.

There is no obvious correlation seen between band depth and solar distance at either 2.95 or 3.1 µm. The phase angle for Eros did not vary beyond the ~30°-50° range (as noted above), and no correlation with band depth is seen with phase angle in this range. The sub-Earth and sub-solar latitudes and longitudes for Eros are known for the times of observation, and similarly no clear correlations are seen with those parameters.

Interestingly, there does appear to be a correlation between band *shape* and solar distance for Eros. Figure 6 shows the 3.1/2.95-µm reflectance ratio vs. solar distance, with a clear decrease in this ratio with increasing distance (and thus with decreasing temperature). This can be interpreted a few ways: it could represent a greater band depth at 3.1 µm at lower temperatures, or it could represent a broadening of the 3-µm band at lower temperatures. This is consistent with an absorber at 3.1 µm being lost at higher temperatures relative to an absorber at 2.95 µm, though again there is little evidence of an overall band depth change. However, we caution that the details of the reflectance ratio trend do not appear to be robust as far as inclusion or exclusion of specific points when we look at band depths instead of reflectance ratios: If all observations are included, the trend is due to a greater 2.95-µm band depth at higher temperatures while the 3.1-µm band depth remains constant. If the lowest-temperature point is excluded, the trend appears due to a *decrease* in 3.1-µm band depth at higher temperatures while the 2.95-µm band depth remains the same. Furthermore, the observing geometry of Eros in June and July 2012 was very similar in sub-observer latitude and longitude, phase angle, and solar distance, yet their band shapes are rather different in Figure 2 and band depths are different in Figure 6. It is possible that Eros passed through a critical temperature threshold during this time period, but we also note that the object was significantly fainter in July 2012, and inspection shows the two spectra are not of similar quality. Finally, as seen on Figure 4, all of the data points for Eros fall within 1-σ of the range defined by the average for Eros and its 1-σ uncertainty, though inspection of Figure 2 shows obvious visual differences in the spectral shapes. Therefore we consider this trend to be intriguing but in need of confirmation.

*4.3 1036 Ganymed*
The 2.95-µm band depth for Ganymed is roughly 2%, though it varies between measurements from 6% to basically zero (Figure 5). This average, and the range, is larger than what is seen for Eros. This would seem to argue against the interpretation of Eros' absorption as endogenic, since no analogs of Ganymed are thought to have hydrated minerals, and this absorption band is almost certainly due to exogenic processes. Therefore, if exogenic processes can create a band depth upward of 3% on Ganymed, we might expect them to be able to make sufficient OH to account for the lesser band depths on Eros.

In the context of the deeper band depth for Ganymed, it is interesting to note the orbital differences between it and Eros. Eros' aphelion is at 1.78 AU, never entering the main asteroid belt and only brushing the inner edge of the Hungaria region. Ganymed, by contrast, traverses the entire asteroid belt and the Cybele region between its perihelion at 1.24 AU and its aphelion in the Hilda region at 4.09 AU. While it has a relatively high inclination, one of its ecliptic crossings is at 2.95 AU in the outer part of the asteroid belt. Thus, it is conceivable that Ganymed may have more opportunity to sweep up exogenic, carbonaceous material than Eros.

As with Eros, the band center differs on different nights, and a band center near 3.1 µm is seen on some of these nights. The trend in the ratio of 3.1-µm reflectance to 2.95-µm reflectance seen for Eros (Figure 6) is also possibly present for Ganymed, with 3.1-µm band depth decreasing at higher temperatures while the 2.95-µm band depth remains constant. As with Eros, we note the existence of this trend in the data but require additional observations to establish it with greater confidence.

*4.4 Other S-class asteroids*
27 observations of 13 S-class asteroids other than Eros and Ganymed have been taken through the LMNOP. Details of these observations are being prepared for publication, but we note that the average 2.95-µm band depth for these observations is 2.2% ± 4.7%, with a median band depth of 1.3%. Eros and Ganymed therefore seem to be consistent with the group of S-class asteroids as a whole, though the uncertainties associated with the group average are consistent with a band depth of zero while individual observations of Ganymed and Eros have small enough uncertainties to establish that absorption bands do exist in at least some orientations.

*4.5 Estimates of water concentration*
We can use data collected by the Dawn spacecraft at Vesta, which includes both gamma-ray and 3-μm spectra, to point toward a more quantitative measurement of hydrogen on Eros and Ganymed using their 3-μm spectra. De Sanctis et al. (2012, 2013) report spectra with band depths of ~1.5-5% at 2.8 μm west of Oppia crater. These band depths are found in regions with hydrogen concentrations of ~250-400 ppm in the GRaND data. The 2.8-μm band depth on Eros and Ganymed is unobservable due to the atmosphere, but in the VIR data a 5% band depth at 2.8 μm results in a ~3.5% band depth at 2.95 μm, similar to what is seen on the NEOs.

A second independent estimate can come from looking at lunar data. A similar semi-quantitative correlation can be made between the hydrogen content as measured by Lawrence et al. (2015) and the 2.95-μm band depth for the same area on the Moon as reported by Clark (2009). Here, we estimate that a ~2% band at 2.95 μm corresponds to ~100 ppm of hydrogen. This is qualitatively consistent with the Vesta-based estimates, and together they suggest, other things being equal, that similar hydrogen concentrations may give rise to the similar band depths on S-class asteroids like Eros and Ganymed, and hydrogen concentrations in the ~few hundred ppm range can be estimated for their surfaces.

A third approach for estimating the amount of hydroxyl on Eros and Ganymed is to follow the approach of Li and Milliken (2014) who used the effective single-particle absorption-thickness (ESPAT) parameter to estimate the amount of hydrated/hydroxylated material in lunar spectra using the band depth at 2.8 μm. Our spectra do not cover that wavelength region, but the spectra of Vesta from (De Sanctis et al. 2012) and the Moon from Sunshine et al. (2009) have reflectances near 2.8 μm that are roughly 0.95-1.0 times their 2.95 μm reflectance. While imprecise, we will use these as analogs for demonstrating consistency. Assuming this range of reflectance ratio, the highest-quality spectra in this dataset result in ESPAT parameter values of ~0.01-0.03 for Eros and Ganymed. The amount of "equivalent water" indicated is a function of the assumed particle size, but using a range of particle bins from 32-53μm to 106-125 μm results in estimated water concentrations from ~30-300 ppm on the asteroids, or ~70-600 ppm of hydrogen. These are, again, consistent with the estimates from other approaches.

For completeness, we mention an additional measurement of hydrogen on Eros for comparison: Peplowski et al. (2015) used data from the NEAR Shoemaker Gamma Ray Spectrometer (GRS) to measure the hydrogen content of Eros, finding a value of $1100^{+1600}_{-700}$ ppm at the spacecraft landing site. This is somewhat higher than the amount we find but can still be considered consistent due to the uncertainties associated each approach and the fact the Peplowski et al. measurement is from a pond area, which may not be representative of Eros' surface as a whole. Indeed, laboratory measurements of hydrogen concentration in L/LL chondrites reported in

Peplowski et al. report an average of 467 ± 464 and 734 ± 754 ppm for L and LL chondrites, respectively, and median values in the ~250-350 ppm range for both meteorite groups, more consistent with our estimates.

We hasten to acknowledge that the relationship between hydrogen content and band depth is nonlinear, and differences in albedo and particle size between Vesta, the Moon, and the NEOs studied here may also be important. Furthermore, neutron spectrometers measure the hydrogen concentration averaged over cm-m scales rather than the 10s of μm scales seen by reflectance spectra, and the relationship between surface absorption band depth and larger-scale average hydrogen concentration may differ between objects. However, even this rough treatment shows that the amount of hydrogen indicated for Eros and Ganymed from their infrared spectra is not unreasonable compared to other similar surfaces. Furthermore, this treatment does not depend upon how the hydrogen was incorporated into the airless surface. This independence is reflected in the consistent estimates from the Moon and Vesta, which are thought to have had different formation mechanisms for its surface hydroxyl.

*4.6 Is the solar wind responsible?*
Lawrence et al. (2015) report hydrogen concentrations up to 120-150 ppm across the lunar highlands. Clark (2009) concluded the 2% absorption band seen in Cassini VIMS observations of the Moon could be due to ~1000 ppm of water. It is generally thought that lunar "water" outside of the polar permanently shadowed regions is created by solar-wind implanted hydrogen into the regolith. Prettyman et al. (2012) argued that the hydrogen concentration on Vesta was much higher than that seen on the Moon and therefore the hydration band seen on Vesta could not be created by solar wind implantation alone. As discussed in the previous section, however, the observations of Eros and Ganymed presented here imply similar hydrogen concentrations as what is seen in the lunar data. As a result, it seems possible that solar wind implantation could have created enough hydroxyl to account for the absorption bands seen on these NEOs.

Does the lack of a trend with temperature or solar distance tell us anything? A simple model might predict an increase in solar wind implantation and OH production, and thus 3-μm band depth, with decreasing distance. Alternately, an equally simple but opposite model might predict OH destruction with smaller solar distance and higher temperatures, with a return of OH with cooler temperatures post-perihelion. We see neither of these trends in these data. However, both of these simple models rely upon changes occurring rapidly on the scale of an asteroid orbit: in the latter case explicitly.

The data are consistent with multiple scenarios for hydroxyl creation. It is possible that it was brought in (and is hosted) by carbonaceous impactors. It is also possible that solar wind implantation and reactions with regolith silicates has allowed the formation of hydroxyl, and that it is at an equilibrium, saturation level. Given the subsolar temperature for Eros and Ganymed in this dataset remains tens of kelvins below what is seen on the hottest parts of the Moon, retention of hydrogen is not unreasonable.

Additional high-quality observations at smaller solar distances and allowing a tighter connection between band depth and specific location will be necessary to further understand the degree to which the solar wind creates hydroxyl on the NEOs.

*5 Summary*
Observations of the two largest NEOs, 433 Eros and 1036 Ganymed taken over their 2011-2013 apparitions, show evidence of absorptions in the 3-µm spectral region consistent with hydroxyl or water. Roughly half of the eight observations of Eros show such absorptions, as do at least seven of the nine observations of Ganymed. These absorptions have band depths in the range of ~1-5% at 2.95 and/or 3.1 µm. Similar band depths on the Moon and Vesta indicate hydrogen amounts of ~100-400 ppm, qualitatively suggesting similar amounts are present on these asteroids. No significant correlation is seen between the 3-µm spectral properties of Eros and Ganymed and other factors save intriguing hints of band shape variation with temperature. Finally, the thermal inertias of these asteroids appear to be ~200 in SI units, consistent with other similar objects.

*Acknowledgments*
ASR gratefully acknowledges consistent support from NSF Planetary Astronomy Program grants 1009710 and 1313144 and from NASA Near Earth Observations grant NNX14AL60G. Contributions by JPE were funded through NASA Solar System Observations grant NN16AE91G and the NASA Spitzer Space Telescope through an award issued by JPL/Caltech.  We acknowledge the sacred nature of Maunakea to many Hawaiians, and our status as guests who have been privileged to work there. Many thanks to the stalwart telescope operators of the IRTF who were instrumental in taking these data through the years, and to Bobby Bus and Eric Volquardsen for developing the "ATRAN part" of the data reduction. Thanks to Carolyn Ernst for useful discussions about the non-intuitive behavior of latitudes and longitudes on irregular small bodies and help with the Small Body Mapping Tool. Reviews by anonymous referees helped strengthen this work, and credit is due to them, whoever they are.

| Date | Objects | Standard Stars | Avg. PW (mm) | Integration time x images |
|---|---|---|---|---|
| 10 Jun 2011 | 1036 | HD 19061, HD 233399 | 5.33 +/- 0.21 | 30 s x 128 |
| 23 Jun 2011 | 1036 | HD 190605 | 5.02 +/- 0.14 | 60 s x 32 |
| 4 Jul 2011 | 1036 | HD 204570 | 2.84 +/- 0.11 | 20 s x 140 |
| 5 Jul 2011 | 1036 | 51 Peg, SAO 107657 | 2.59 +/- 0.08 | 10 s x 24 |
| 27 Aug 2011 | 1036 433 | HD 222794, HD 23169 | 0.83 +/- 0.05 | 60 s x36 60 s x44 |
| 28 Aug 2011 | 1036 | HD 19061, HD 217429, HD 7983 | 1.27 +/- 0.09 | 60 s x 8 |
| 30 Oct 2011 | 1036 433 | HD 19061, HD 233399 | 1.87 +/- 0.17 | 45 s x 16 45 s x 34 |
| 6 Dec 2011 | 1036 433 | SAO 129922 Hya 64, SAO 60387 | 3.19 +/- 0.50 | 10 s x 8 6 s x 8 |
| 25 Jan 2012 | 433 | HD 107146 | 0.62 +/- 0.01 | 40 s x 12 |
| 30 Jan 2012 | 1036 | HD 12846, HD 60298 | 1.26 +/- 0.08 | 60 s x 28 |
| 1 May 2012 | 433 | HD 101060, HD 91398 | 0.48 +/- 0.2 | 45 s x 32 |
| 4 Jun 2012 | 433 | HD 105328 | 1.88 +/- 0.06 | 45s x 32 |
| 3 Jul 2012 | 433 | HD 117286 | 0.99 +/- 0.05 | 45 s x 24 |

**Table 1:** Details of Observing Nights. The last column shows the number of images along with the integration time per image. The precipitable water (PW) column is calculated from the average value fitted for each separate asteroid and standard star observation.

| UT Date | Midtime (UT) | V Mag | Solar Dist (AU) | Earth Dist (AU) | Phase Angle | SE Lat | SE Lon | SS Lat | SS Lon |
|---|---|---|---|---|---|---|---|---|---|
| 27 Aug 2011 | 14:57 | 13.27 | 1.48 | 1.04 | 43.1° | -85° | 24° | -90° | 124° |
| 30 Oct 2011 | 12:11, 12:54 | 11.6 | 1.27 | 0.53 | 47.5° | -49° | 347°, 35° | -85° | 8°, 57° |
| 6 Dec 2011 | 14:27 | 10.4 | 1.18 | 0.33 | 47.6° | +71° | 314° | -76° | 342° |
| 25 Jan 2012 | 11:50 | 8.64 | 1.13 | 0.18 | 31.2° | +86° | 53° | +76° | 53° |
| 1 May 2012 | 5:23 | 11.4 | 1.35 | 0.50 | 38.8° | +86° | 274° | +89° | 57° |
| 4 Jun 2012 | 6:07 | 12.5 | 1.46 | 0.76 | 40.7° | +88° | 262° | +87° | 39° |
| 3 Jul 2012 | 5:53 | 13.28 | 1.55 | 1.02 | 40.5° | +89° | 357° | +84° | 42° |

**Table 2:** Observing Circumstances for 433 Eros. In this and the following table, "SE Lat" and "SS Lat" are the sub-Earth and sub-solar latitude, respectively, at the time of the observation, with "SE Lon" and "SS Lon" the sub-Earth and sub-solar longitudes. These values are taken from the JPL Horizons ephemeris generator.

| UT Date | Midtime (UT) | V Mag | Solar Dist (AU) | Earth Dist (AU) | Phase Angle | SE Lat | SE Lon | SS Lat | SS Lon |
|---|---|---|---|---|---|---|---|---|---|
| 10 Jun 2011 | 14:00 | 11.45 | 1.54 | 0.89 | 39.0° | +44° | 213° | +18° | 297° |
| 23 Jun 2011 | 14:28 | 11.20 | 1.46 | 0.81 | 41.9° | +58° | 52° | +21° | 44° |
| 4 Jul 2011 | 11:54 | 11.02 | 1.41 | 0.75 | 44.5° | +65° | 176° | +24° | 170° |
| 5 Jul 2011 | 15:07 | 11.01 | 1.40 | 0.75 | 44.8° | +66° | 46° | +25° | 40° |
| 27 Aug 2011 | 13:12 | 10.18 | 1.24 | 0.53 | 52.3° | +74° | 112° | +32° | 95° |
| 28 Aug 2011 | 13:45 | 10.15 | 1.24 | 0.52 | 52.2° | +74° | 250° | +32° | 232° |
| 30 Oct 2011 | 10:25 | 8.37 | 1.40 | 0.40 | 1.4° | +17° | 336° | +17° | 336° |
| 6 Dec 2011 | 7:48 | 10.9 | 1.61 | 0.76 | 26.4° | -4° | 272° | +6° | 283° |
| 30 Jan 2012 | 6:04 | 13.0 | 1.97 | 1.59 | 13.0° | -7° | 204° | -7° | 220° |

**Table 3:** Observing Circumstances for 1036 Ganymed. The column headings are the same as the previous table. The longitude is calculated with the prime meridian defined as the sub-solar longitude at 0:00 UT on 10 June 2011.

| Date | η | Tss (K) | 2.95 BD | 3.1 BD |
|---|---|---|---|---|
| 27 Aug 2011 | 0.89 | 329 | 2.2 ± 1.4% | 3.8 ± 1.4% |
| 30 Oct 2011 | 1.03, 1.08 | 342, 338 | 0.7 ± 1.9%, 4.4 ± 1.1% | 1.0 ± 1.7%, 4.0 ± 1.1% |
| 6 Dec 2011 | 1.14 | 346 | 2.9 ± 1.3% | 0.7 ± 1.2% |
| 25 Jan 2012 | 1.12 | 355 | 2.2 ± 0.7% | 1.6 ± 0.7% |
| 1 May 2012 | 0.87 | 346 | 2.9 ± 3.0% | 1.9 ± 3.1% |
| 4 Jun 2012 | 0.92 | 329 | 5.0 ± 2.7% | 7.7 ± 2.6% |
| 3 Jul 2012 | 0.99 | 313 | 0 ± 2.0% | 4.0 ± 1.9% |

**Table 4:** Results for Thermal Modeling and Band Depths for 433 Eros. The beaming parameter is chosen by the procedures listed in Section 3.2, and Tss is the implied subsolar temperature. The band depths are calculated using averages over the 0.01 μm ranges around the wavelength listed, measured from a straight-line continuum between 2.45 and 3.4 μm. Negative band depths are reported as zero. The uncertainty is the 1-σ standard deviation, with uncertainties in the continuum included.

| Date | η | Tss (K) | 2.95 BD | 3.1 BD |
|---|---|---|---|---|
| 10 Jun 2011 | 1.03 | 312 | 3.7 ± 0.5% | 3.1 ± 0.5% |
| 23 Jun 2011 | 0.98 | 324 | 1.7 ± 0.6% | 2.6 ± 0.6% |
| 4 Jul 2011 | 0.96 | 331 | 1.0 ± 1.8% | 0.1 ± 1.8% |
| 5 Jul 2011 | 0.98 | 331 | 0 ± 1.5% | 1.4 ± 1.4% |
| 27 Aug 2011 | 0.97 | 352 | 2.2 ± 0.8% | 1.2 ± 0.7% |
| 28 Aug 2011 | 0.94 | 355 | 2.8 ± 1.3% | 1.4 ± 1.3% |
| 30 Oct 2011 | 0.85 | 342 | 1.6 ± 1.1% | 1.2 ± 1.2% |
| 6 Dec 2011 | 0.91 | 314 | 6.2 ± 2.1% | 0 ± 2.0% |
| 30 Jan 2012 | 0.90 | 285 | 0 ± 2.3% | 1.7 ± 2.2% |

**Table 5:** Same as Table 4, but for 1036 Ganymed.

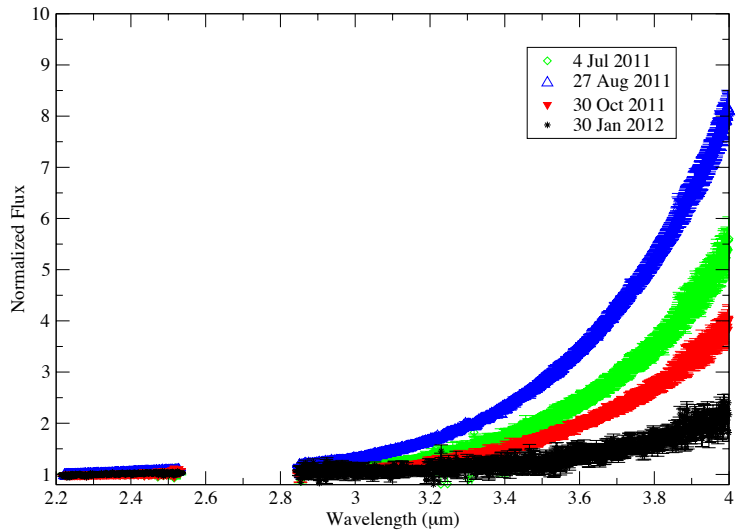

**Figure 1:** Four observations of 1036 Ganymed at different solar distances show the contribution of thermal flux to the overall asteroidal flux. The warmest measurement of Ganymed shown, on 27 August 2011, occurred at a solar distance of 1.24 AU and a calculated Tss of 352 K, while the measurement on 30 Jan 2012 occurred at a solar distance of 1.97 AU and calculated Tss of 285 K. This temperature difference leads to a factor of 4 difference in thermal flux at 4 μm. The effect of phase angle can also be seen here: while the 4 Jul and 30 October observations were taken at similar solar distance, their phase angles were 45° and 1°, respectively. The smaller phase coefficient of emitted radiation compared to reflected light means that the higher phase angle observations have a larger emitted component relative to the reflected component.

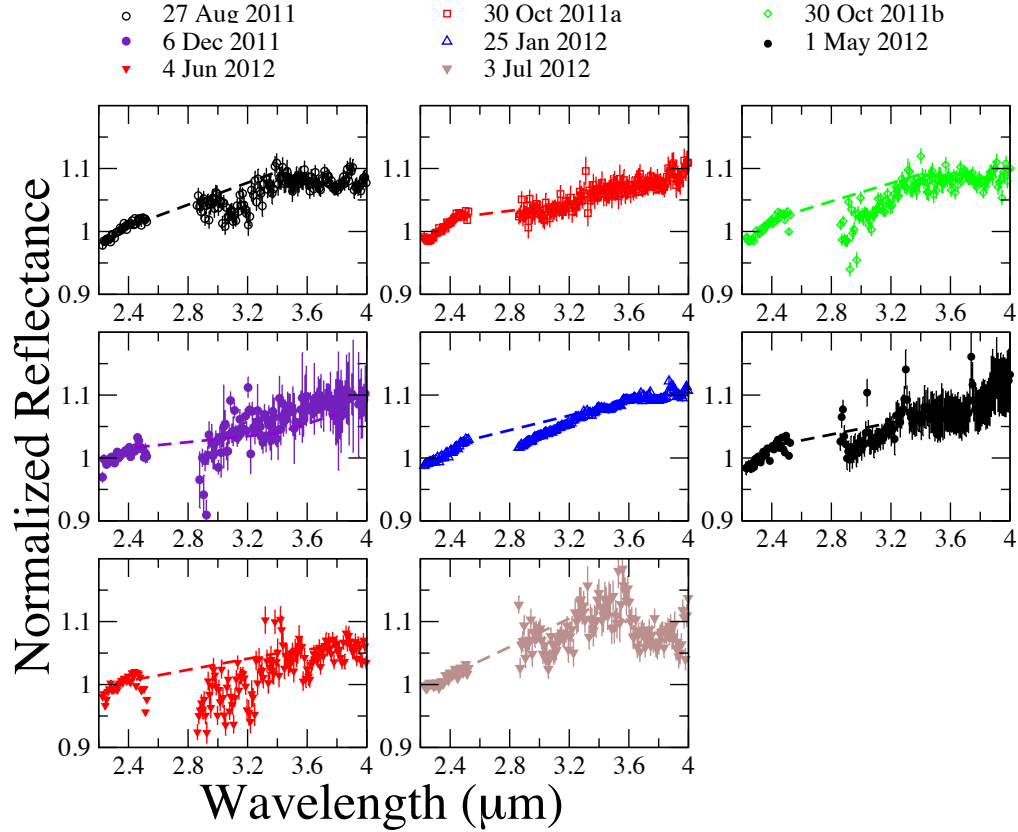

**Figure 2:** Spectra of 433 Eros, arranged in order of observational date. All spectra are normalized to 1 at 2.35 µm. Dashed lines show a straight-line continuum drawn from 2.45 µm to 3.40 µm. Eros was visited twice on 30 October 2011, and the spectra are reported separately here. Many of the Eros spectra show only limited evidence for absorption in the 3-µm region. However, spectra from several nights, notably 27 August 2011, 30 Oct 2011b, 25 Jan 2012, and 1 May 2012, do appear to have absorption bands. Some spectra appear to have minima in the 3.1-3.2 µm region, but others appear to have spectra more reminiscent of what is seen in common low-albedo asteroids and carbonaceous chondrites.

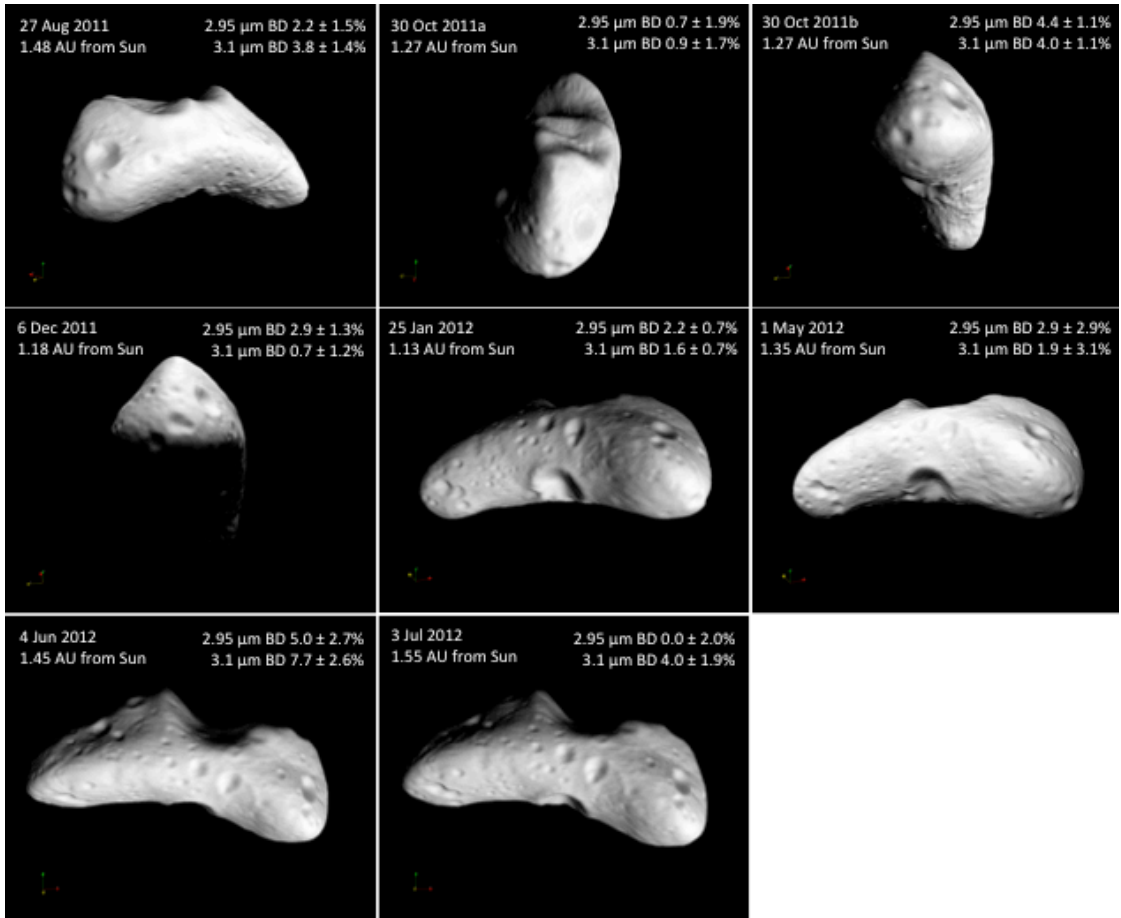

**Figure 3:** Visualizations of Eros at the midtimes of observing, from the Small Body Mapping Tool (Kahn et al. 2011). The panels are arranged in the same way as the spectra in Figure 2, to allow easy comparison. Also included are the date, solar distance, and 2.95- and 3.1-μm band depths and uncertainties. Taken together, the band depths and visualizations appear potentially consistent with some variation on Eros' surface, although visual inspection shows that the spectra vary in quality and, for instance, poorer-quality data is likely the cause of the different band depths for 4 June 2012 vs. 3 July 2012 observations with the same viewing geometry.

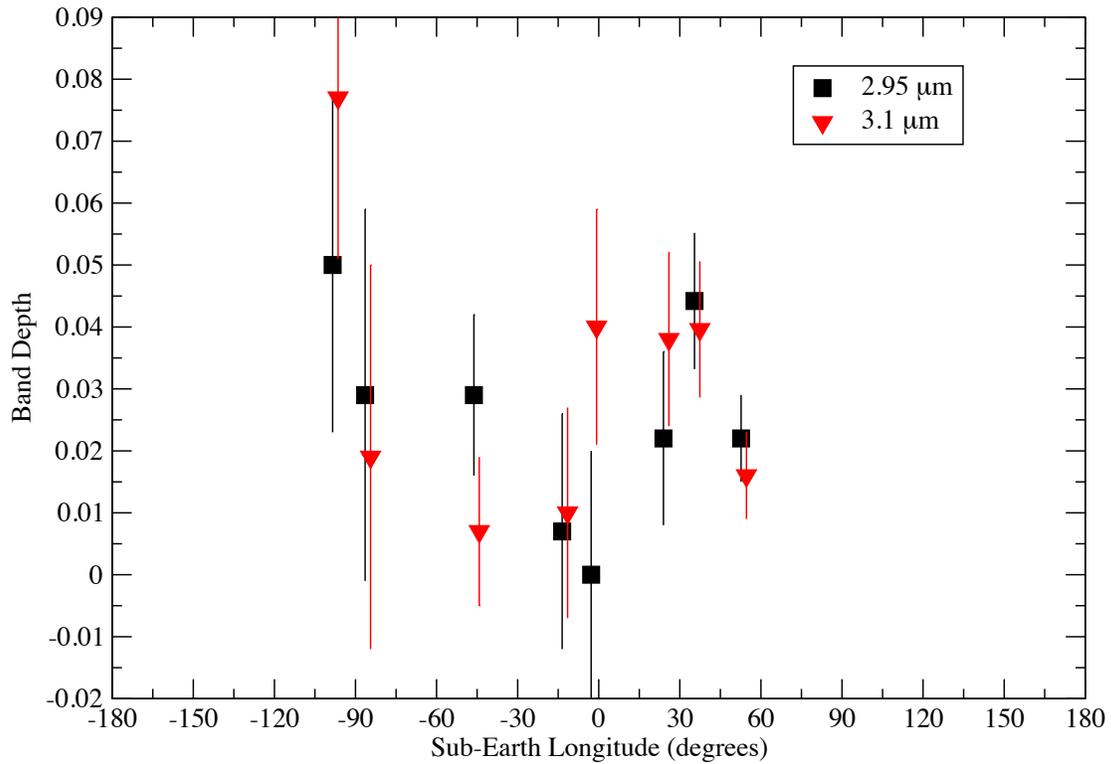

**Figure 4**: The 2.95- and 3.1-μm band depths on Eros are potentially correlated with surface features. Taken at face value, the 2.95-μm band depth seems lower near the prime meridian and constant elsewhere, while the 3.1-μm band depth seems lower across most of Eros' Himeros-Shoemaker hemisphere. While intriguing, the relatively large uncertainties require further work before a conclusion can be reached. The 2.95-μm band depths are presented at their sub-Earth longitude at the mid-time of observing, while the 3.1-μm band depths are offset from the 2.95-μm band depths by 2 degrees of longitude for clarity.

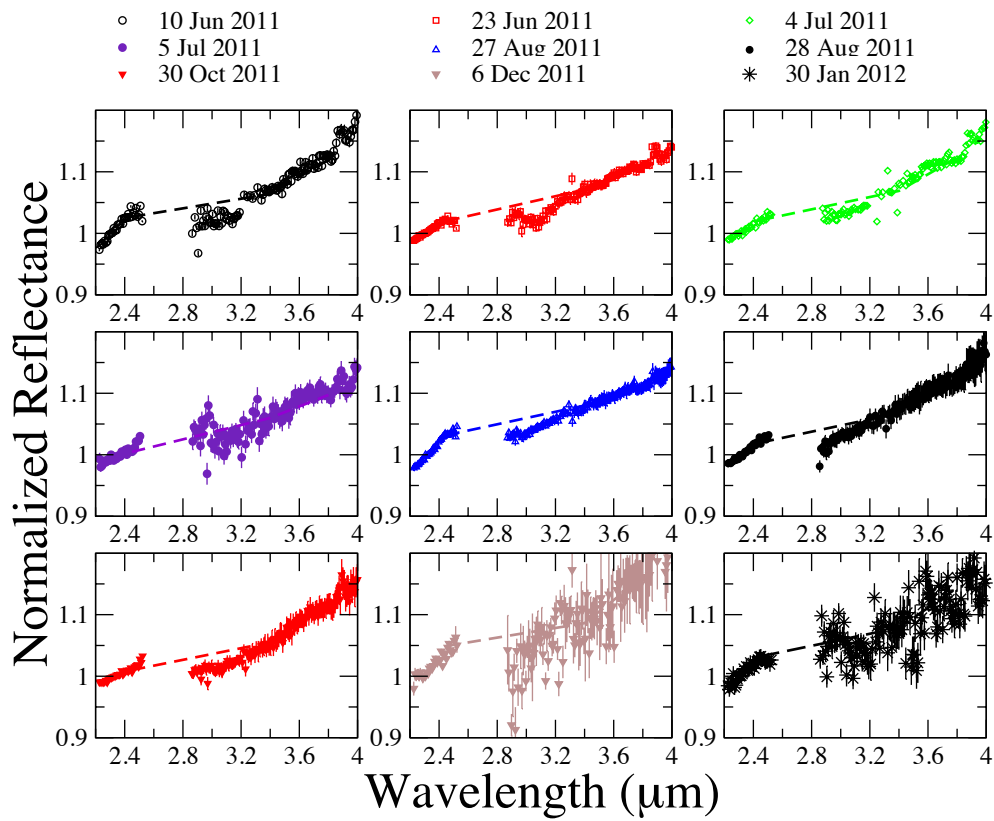

**Figure 5:** Spectra of 1036 Ganymed, arranged, normalized, and offset as in the previous figure. Dashed lines show a straight-line continuum drawn from 2.45 µm to 3.40 µm. All Ganymed spectra show evidence of an absorption band in the 3-µm region, though the band shapes and depths differ. As with Eros, some spectra appear to have minima in the 3.1-3.2 µm region, but others appear to have spectra more reminiscent of what is seen in common low-albedo asteroids and carbonaceous chondrites.

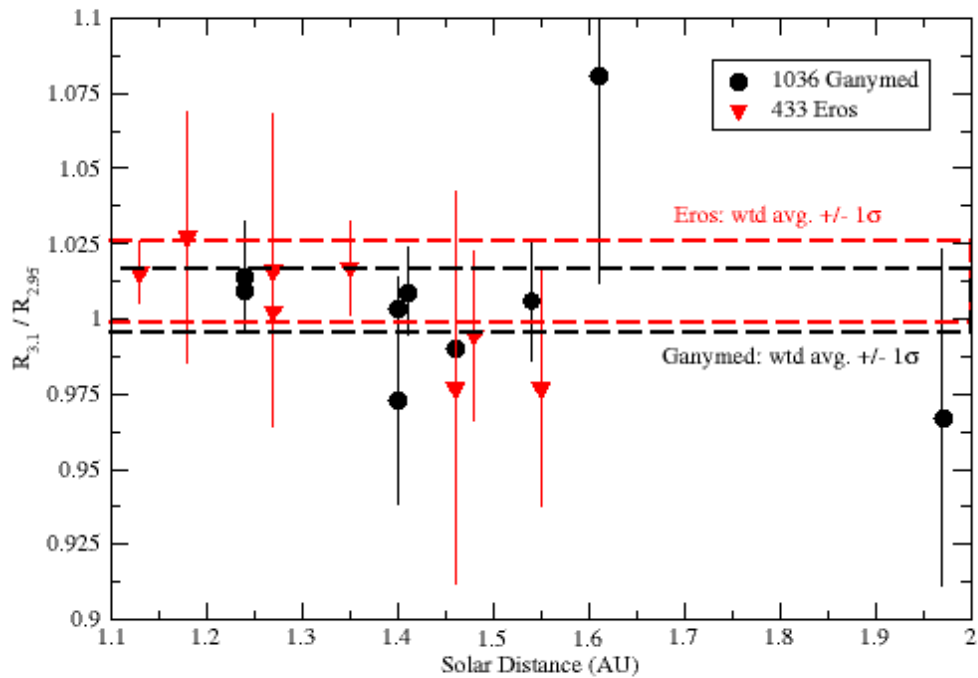

**Figure 6:** The ratio of the 3.1-μm reflectance to the 2.95-μm reflectance is a measure of the 3-μm band shape. Points and lines associated with Eros are in red, those with Ganymed are in black. Both Eros and Ganymed show a general decrease in this ratio with increasing semi-major axis (and thus with decreasing temperature). While this trend is intriguing, it is highly dependent on the presence or absence of specific data points. We also note that all measurements (including their 1-σ uncertainty) in the apparent trend for a given asteroid fall within the 1-σ uncertainty of the weighted average for that asteroid (dashed lines). We cannot conclude whether this trend will ultimately prove significant, but consider it worth further study.